\def\figref#1{Fig.\ \ref{#1}} 

\newcommand{\beq}{\begin{equation}} \newcommand{\eeq}{\end{equation}}
\def\note#1{\def\dash{\hbox{\rm---}}{\bf~[[}~{\tt #1}~{\bf]]~}}
\def\ltsim{\hbox{\kern.25em\raise.5ex\hbox{$<$}\kern-.75em\lower.5ex
   \hbox{$\sim$}\kern.25em}}
\documentstyle[preprint,aps,epsfig]{revtex}
\begin{document}            \draft  \tightenlines 
\title{
Slow relaxation, confinement, and solitons} 
\author{
L. S. Schulman,$^1$\footnote{Email: schulman@clarkson.edu} E. Mih\'okov\'a,$^2$
A. Scardicchio,$^3$ P. Facchi,$^3$ M. Nikl,$^2$ K. Pol\'ak,$^2$ and B.~Gaveau,$^4$}
\address{
$^1$ Physics Department, Clarkson University, Potsdam, New York 13699-5820, USA\\
$^2$Inst.\ of Physics, Acad.\ of Sciences of the Czech Rep.,
Cukrovarnick\'a~10, 162~53~Prague~6, Czech~Republic\\
$^3$Dipartimento di Fisica, Universit\`a di Bari  I-70126 Bari, Italy\\
$^4$ Universit\'e Paris 6, Case courrier 172, 4 Place Jussieu, 75252 Paris C\'edex 05, France}
\date{\today} \maketitle
\begin{abstract}Millisecond crystal relaxation has been used to explain anomalous decay in doped alkali halides. We attribute this slowness to Fermi-Pasta-Ulam solitons. Our model exhibits confinement of mechanical energy released by excitation. Extending the model to long times is justified by its relation to solitons, excitations previously proposed to occur in alkali halides. Soliton damping and observation are also discussed.
\end{abstract}

\bigskip


\narrowtext

Crystals respond in picoseconds. This is the characteristic time for all sorts of lattice adjustments, for example passage to a ``relaxed excited state." Nevertheless, an anomaly in the decay of luminescence in doped alkali halides has led us to conclude \cite{decaykinetics,anomalousall} that relaxation can sometimes take place on a scale of milliseconds. It's as if the tides moved like tectonic plates. In this article we propose that the formation and slow decay of Fermi-Pasta-Ulam (FPU) solitons \cite{fermi,ford,tabor} can account for this dramatic slowdown. The presence of FPU solitons in alkali halides has been previously proposed \cite{sievers,bickham} and we believe that particular features of our system \cite{decaykinetics,anomalousall} enable their production and observation.

We begin with an overview. KBr (say) is dilutely doped with Pb$^{2+}$. One can think of the Pb and its 6 surrounding Br's as a quasimolecule. This quasimolecule is excited by a flash of UV light and because of the Jahn-Teller effect distorts considerably. In the absence of the constraining lattice one would expect expansion along one of the axes of perhaps 15\%. The distortion is asymmetric (there is shrinking along other axes) and the Jahn-Teller interaction causes well-understood degeneracy-breaking effects. It is the subsequent return to the ground state that provides evidence for slow lattice relaxation. The excited quasimolecule can go to a {\em radiative} level with a 25~ns lifetime or to a {\em metastable} level with an 8~ms lifetime. The luminescence decay however is {\em not\/} the sum of two exponentials, but rather, after the initial burst from the fast level, molecules in the metastable level exhibit an enhanced decay rate, which we have explained to be due to lattice-induced coupling to the fast level. This continues for milliseconds and implies that the lattice itself takes ms to accommodate to the distortion. This is the slow relaxation, and our ability to use this assumption to fit a variety of systems in a large range of temperatures is the evidence for this phenomenon. 

Because the distortion of the quasimolecule is asymmetric, a reasonable model is a linear diatomic chain in which a particle on one end is subject to a strong push. (This is the Br adjacent to the Pb.) Nonlinearity enters because of the large displacements. Moreover, to account for the 3-dimensional environment we add a potential (simulating off-chain neighbors) that tends to return each ion to its normal lattice position. This model displays a remarkable property: reverberations in the chain due to the push are confined. Oscillations occur only in the 2 to 4 neighbors of the first ion. Except for a small initial pulse traveling ballistically, no energy escapes. These assertions are based on numerical integration for $\ltsim500$ lattice time units. Accurate integration of even the 1-dim classical case for $10^9$ units ($\sim\mskip 1mu$ms) appears out of reach, so if we claim this confinement underlies slow relaxation, more general arguments are needed.

As we show below, the confinement reflects the creation of a FPU soliton. This structure lives {\em almost} forever (``almost" refers to Arnold diffusion). We thus go from the problem of {\em fast\/} relaxation to that of {\em no} relaxation. To explain the ultimate decay of the soliton we note the apparent coincidence that the relaxation is on the same scale as the electromagnetic decay of the metastable level. This suggests that the electronic degrees of freedom draw energy from the soliton. This damping yields patterns of lattice relaxation that follow essentially the time dependence used in explaining the experiments. We also found (numerically) that even after the quasimolecule decays the soliton remains. 

This explanation of slow relaxation also leads to the question of direct observation. If indeed for $\sim\mskip 1mu$5~ms small bunches of atoms vibrate at a frequency above the acoustical phonon branch, what probe could reveal this?

{\em Details of the model.}~ Following the UV pulse, and faster than any scale considered here, the electronic wave function, $\psi$, distorts. The quasimolecule is pushed hard along one axis and shrunk along the others. We focus on a ray of atoms along the expansion axis. The first atom is subjected to a large force which it transmits to the others. We simplify the influence of $\psi$ by pretending there is a fictitious zeroth particle displaced by a fixed amount from its equilibrium. To allow the use of a 1-dimensional chain we supply each atom on the chain with a ``neighbor"-force that attracts the atom to its nominal place in the lattice. Note that the asymmetric stresses imply lesser deformation for off-axis atoms. The interatomic potential is taken to be $V(u)= M\omega_0^2 \left(u^2 + \lambda u^4\right)/2$. The Hamiltonian is
\beq
H=\sum_{n=1}^N   \left\{\frac{P_n^2}{2M}+\frac{rp_n^2}{2M}
  + V(Q_n-q_n)) +V(Q_n-q_{n-1})+\nu \left(V(Q_n^2)+V(q_n^2)\right) \right\} \;,
\label{hamiltonian}
\eeq
where $\nu$ is the effective number of neighbors. $Q_1$ is the first atom to the right of the Pb, followed by $q_1$, $Q_2$, etc. The $Q$-particles have mass $M$, the $q$s, $M/r$---but we immediately rescale so that $M=1$ and $\omega_0=1$. For KBr, $r\approx2$. The impact of $\psi$ is expressed by setting (the non-dynamic) $q_0\neq0$.

The system was solved classically by numerical integration. All positions and momenta were initially zero, with $q_0$ providing the driving force. \figref{histories} shows runs with $\nu=0$ and 4 ($\lambda=1$). In the first case there is wave propagation, but with $\nu=4$ the energy is confined. A small pulse leaves the system initially, but afterward only the first atoms oscillate, mostly just the first two. \figref{linear} shows the positions for $\nu=4$ but $\lambda=0$. Evidently there is no confinement. These results are insensitive to the exact parameters and persist both for longer times and for more particles. They become more dramatic as $r$ moves away from unity, consistent with the trend in anomalous decay data \cite{decaykinetics,anomalousall}.

\note{\figref{histories} and \figref{linear} belong here.}

We next study the frequency spectrum of the confined vibrations. To connect to solitons this will be related to lattice phonons properties. When $\nu=4$, $r=2$, $N=15$, the phonon spectrum has a gap between the acoustical and optical branches running from 2.45 to 3.46 ($\omega_0=1$). In \figref{intensities} we plot the intensity of the Fourier transform of $q_1(t)$ for a period of 200 time units for three values of $\lambda$. The curve peaked below $\sim$~2.45 is the linear case, $\lambda=0$. The next two are $\lambda=0.5$ and~1. They are dominated by their peaks, which come at 2.86 and 3.27 respectively---within the gap. (The bump just below 4 is part of the $\lambda=1$ spectrum; there is a smaller one above 2.86 for the $\lambda=0.5$ case.) If $q_0$ is varied, the peak-locations change, consistent with scaling.

\note{\figref{intensities} belongs here.}

Besides the extreme spatial localization, another property of our confinement curves---perhaps the characteristic soliton feature in the FPU study---is the failure of the energy to disperse among the normal modes. In \figref{modeanal} we show the energy dispersal for various times. Unlike FPU it is never concentrated in one mode, because our initial conditions already have dispersion. Nevertheless, it is clear that 
while energy can shift from mode to mode (cf.\ time-25 in the figure), it tends to return (repeatedly). Note that we show only quadratic contributions to energy, so the sum displayed need not be conserved.

\note{\figref{modeanal} belongs here.}

An extensive recent literature on FPU solitons exists \cite{sievers,bickham,chubykalo,burlakov,others}, and their presence in alkali halide crystals has been suggested \cite{sievers,bickham}. Like our confinement mode, these excitations are dominated by a single frequency. In \cite{burlakov} the {\em diatomic} chain is studied, and the soliton frequency is found between the acoustical and optical phonon bands.

The ability of our systems \cite{decaykinetics,anomalousall} to support solitons can be attributed to several features. 1.) The distortion (\`a la Jahn-Teller) makes the system closer to 1-d than 3-d, bearing in mind \cite{bickham} the lesser tendency to form solitons in 3-d. 2.) Considerable enhancement of all soliton effects develops as $|r-1|\uparrow$. 3.) From numerical work we find that the boundary condition at the impurity site, based on our physical picture, enhances localization.

{\em Decay of the soliton.}~ Given the confinement--soliton connection, why does the crystal relax at all? It is a common misapprehension that the FPU soliton lives forever; it does not. In \cite{ford} it is shown that FPU dynamics is equivalent to Henon-Heiles chaos. For our parameters, however, as for FPU, the excitation is extremely long-lived, disappearing only because of Arnold diffusion. However, even if for KBr:Pb$^{2+}$ this time scale would be ms, its sensitivity to parameter changes imply that there is no reason to expect the same for other doped alkali halides with ms anomalous decay. One should also realize that the soliton is robust. Defects may shift its properties but not the fact of its existence. Moreover, anomalous decay has been seen with various levels of crystal defects. For these reasons we turn to other degrees of freedom, and in view of the apparent coincidence of time scales, it is the electronic coupling that we argue provides decay.

The decay of the metastable A$_{1u}$ level ($\psi_u$) to the A$_{1g}$ ground state ($\psi_g$) is forbidden by electromagnetic selection rules. The way the decay {\em does} occur is by coupling $\psi_u$ to vibronic modes, which in turn couple to $\psi_g$. We invoke almost the same matrix elements, but in a different order: from the solitons to $\psi_u$ to other lattice vibrations. In one case one has $\langle\psi_u |H| \hbox{vib.}\rangle \allowbreak \langle\hbox{vib.} |H| \psi_g\rangle$, in the other $\langle\hbox{vib.} |H| \psi_u\rangle \allowbreak \langle\psi_u |H| \hbox{vib.}\rangle$. Density of states factors may differ. Nevertheless, it is plausible that the rates are comparable, and this is our assumption.

To incorporate this in our classical model we view the coupling as a damping: energy in the solitons is converted to undifferentiated vibrational energy via electronic coupling. This affects primarily the Br in the quasimolecule, so that classically we included $-\gamma \dot Q_1$ (with the physical $\gamma^{-1}\sim$ ms). In \figref{expdamp} we show a smoothing (average) of the oscillations of $Q_1$ over an extended time period with damping. This is also superimposed on a curve $1-\exp(-\Gamma t)$, with $\gamma\approx\Gamma$. The reason for this comparison is that $1-\exp(-\Gamma t)$ is one of the relaxation forms adopted in \cite{anomalousall} in the slow-relaxation explanation of anomalous decay. We thus recover the desired form for relaxation \cite{delay}.

\note{\figref{expdamp} belongs here.}

\noindent{\em Remarks.} 1.) We expect the quantum version our theory to give similar results. In \cite{sievers} is a comment on quantum treatments, and we remark that the quantization of the soliton can be accomplished by path integral WKB methods \cite{pathintegral}. This is because the phase space trajectory of the soliton is essentially a torus. Weak phonon coupling provides damping, as in \cite{caldeira}, and a lifetime. \hspace{2pt} 2.) For nonzero temperature other mechanisms of soliton decay enter and indeed our data fits \cite{anomalousall} indicate increased ``$\Gamma$" with temperature. Mechanisms such as those studied in \cite{lindenberg} should play a role, although as observed there, the effect may be sensitive to details of the system.

{\em Detection of solitons.}~ Our explanation of slow relaxation implies that for periods $\sim$5~ms subsequent to a UV pulse, small bunches of atoms (perhaps 5 on a side) oscillate at frequencies above the acoustical phonon band. For KBr this is at roughly $\nu_a= 3\!\times\!10^{12}\mskip 1 mu$s$^{-1}$ \cite{karo}. A direct way to see these would be by off-resonant Raman scattering, to avoid the difficulties of strong luminescence. The crystal would simultaneously be illuminated with a Raman laser and, for soliton creation, UV light (e.g., a hydrogen lamp). One would then look for new peaks, in addition to the characteristic peaks of (PbBr)$_6$ and the second order Raman spectrum of KBr. At liquid helium temperatures we expect Raman lines above the acoustical branch and therefore any new peak is expected to be near the laser excitation line. When the UV light is switched off (terminating A-band excitation of Pb$^{2+}$) these Raman lines should disappear. We also anticipate requiring a higher Pb concentration than is now used in decay experiments.

Finally we observe that the soliton frequency corresponds to an energy, $\hbar\omega\sim10\mskip 2mu$meV. The energy dumped into vibrational modes is much larger (perhaps as much as 650$\mskip 2mu$meV), so that the semiclassical approach to the study of the soliton is reasonable.

\acknowledgments
We thank V. \v C\'apek and S. Pascazio for helpful discussions. This work was supported by NSF grants PHY 97 21459 and PHY 00 99471 and by the Czech grant ME382.


\def\sqmeas{7.5cm}\def\precap{5pt}
\begin{figure}[t]
\centerline{\epsfig{file=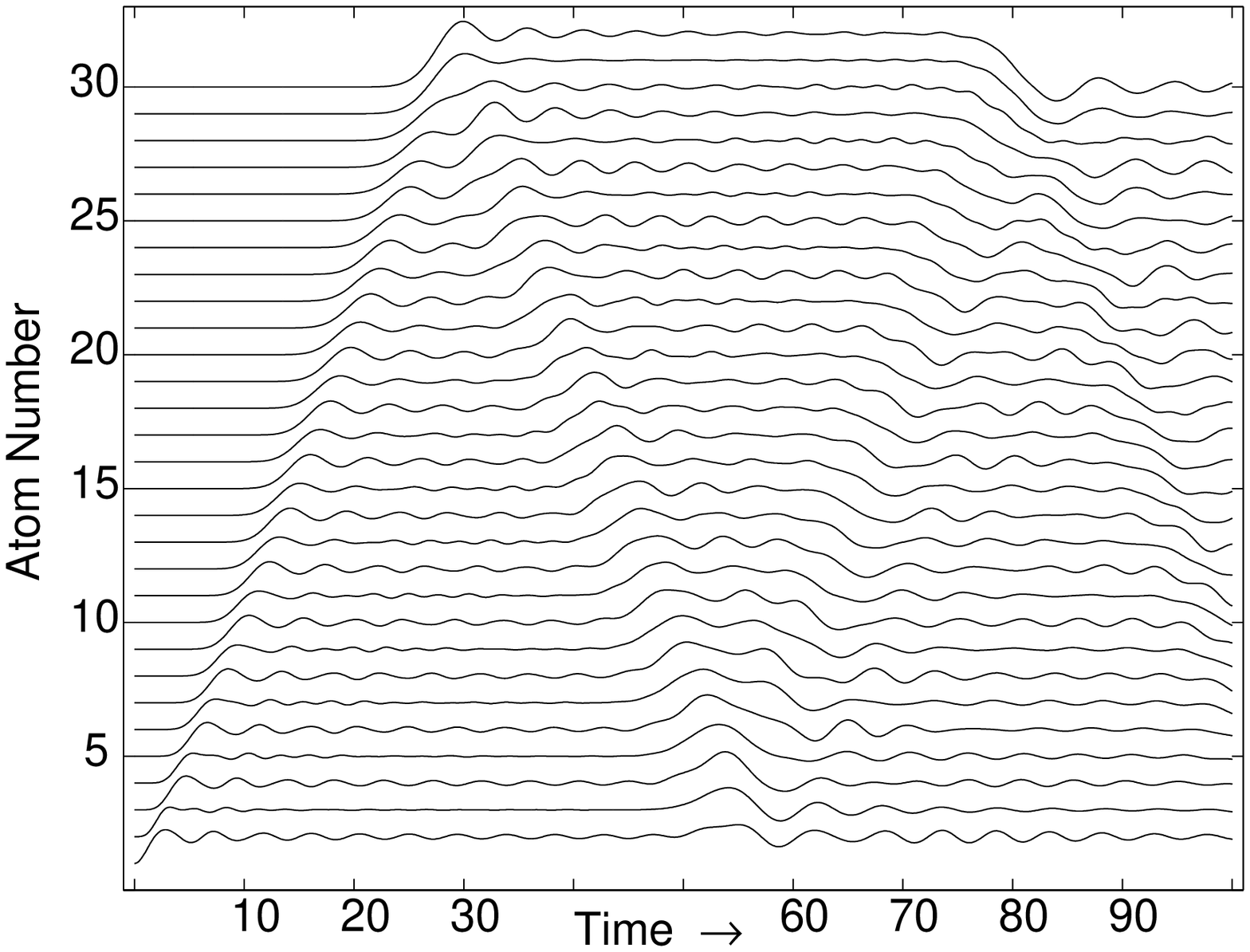,height=\sqmeas,width=\sqmeas}
            \epsfig{file=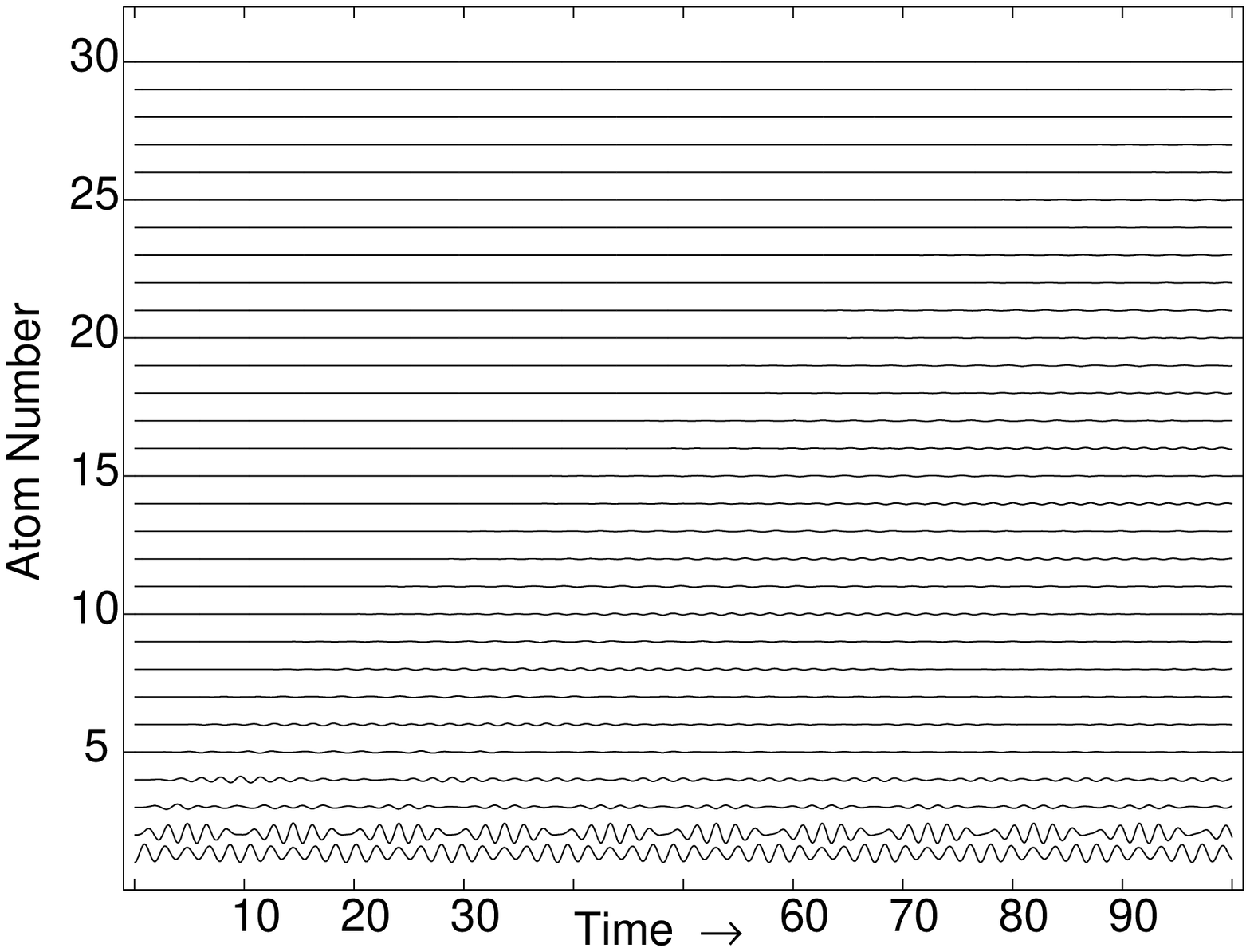,height=\sqmeas,width=\sqmeas}} \vskip \precap
\caption{Position vs.\ time. The unconfined history has $\nu=0$, no force from off-axis atoms. There is wave propagation. The reflection is an artifact of the boundary. The case $\nu=4$ gives confinement (for which the boundary is irrelevant). The lowest curve is $Q_1(t)$; above it is $q_1(t)$, etc. Common parameters are $\lambda=1$, $q_0=1$, $N=15$ (30 atoms), $r=2$.}
\label{histories}\end{figure}
\begin{figure}[t]
\centerline{\epsfig{file=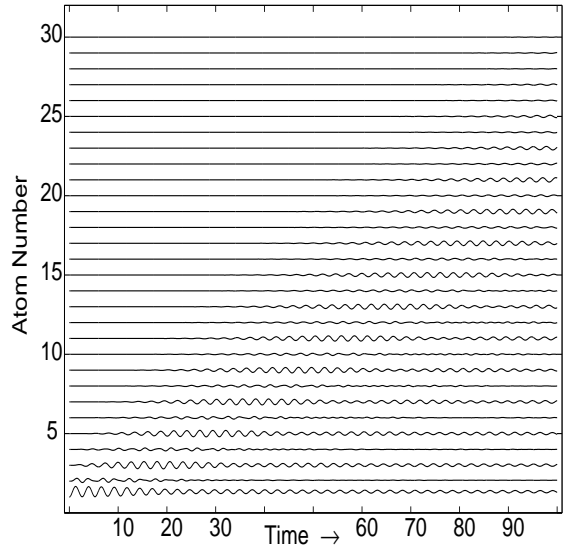,height=\sqmeas,width=\sqmeas}} \vskip \precap
\caption{Position vs.\ time for a system that with nonlinearity would show confinement. Parameters as in \figref{histories} ($\nu=4$), but $\lambda=0$.}
\label{linear}\end{figure}
\def\ymeas{6.5cm} \def\xmeas{7.5cm} \def\precap{5pt}
\begin{figure}[t]
\centerline{\epsfig{file=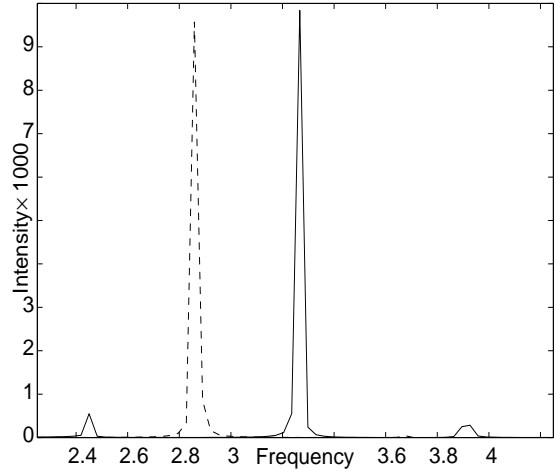,height=\ymeas,width=\xmeas}} \vskip \precap
\caption{
Fourier transform (intensity) of $Q_1(t)$ for various levels of nonlinearity. The dashed curve is $\lambda=0.5$. The two solid-line curves do not overlap, with amplitude to the left of 2.45 corresponding to $\lambda=0$ and to the right of 3, to $\lambda=1$. For $\lambda=0$, the peak is just within the acoustical band. With $\lambda=0.5$ (dashed line) almost all energy is in the peak, which now lies well within the gap. Finally, with $\lambda=1$ (solid line, again) the peak has almost reached the bottom of the optical branch and a second small peak beyond the optical spectrum is more pronounced.}
\label{intensities}
\end{figure}
\def\ymeas{6cm} \def\xmeas{7.5cm} \def\precap{5pt}
\begin{figure}[t]
\centerline{\epsfig{file=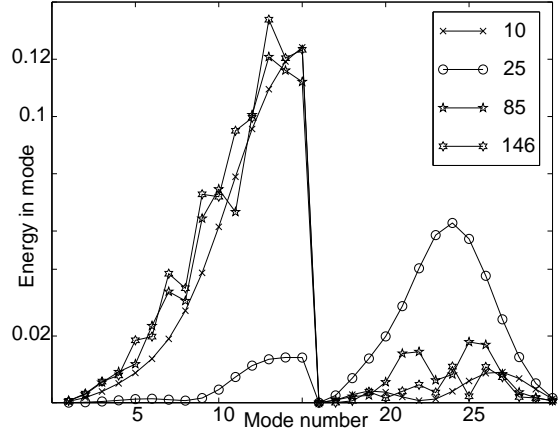,height=\ymeas,width=\xmeas}} \vskip \precap
\caption{Energy per normal mode at various times. Parameters as in \figref{histories}, $\nu=4$.}
\label{modeanal}\end{figure}
\def\ymeas{6cm} \def\xmeas{7.5cm} \def\precap{5pt}
\begin{figure}[t]
\centerline{\epsfig{file=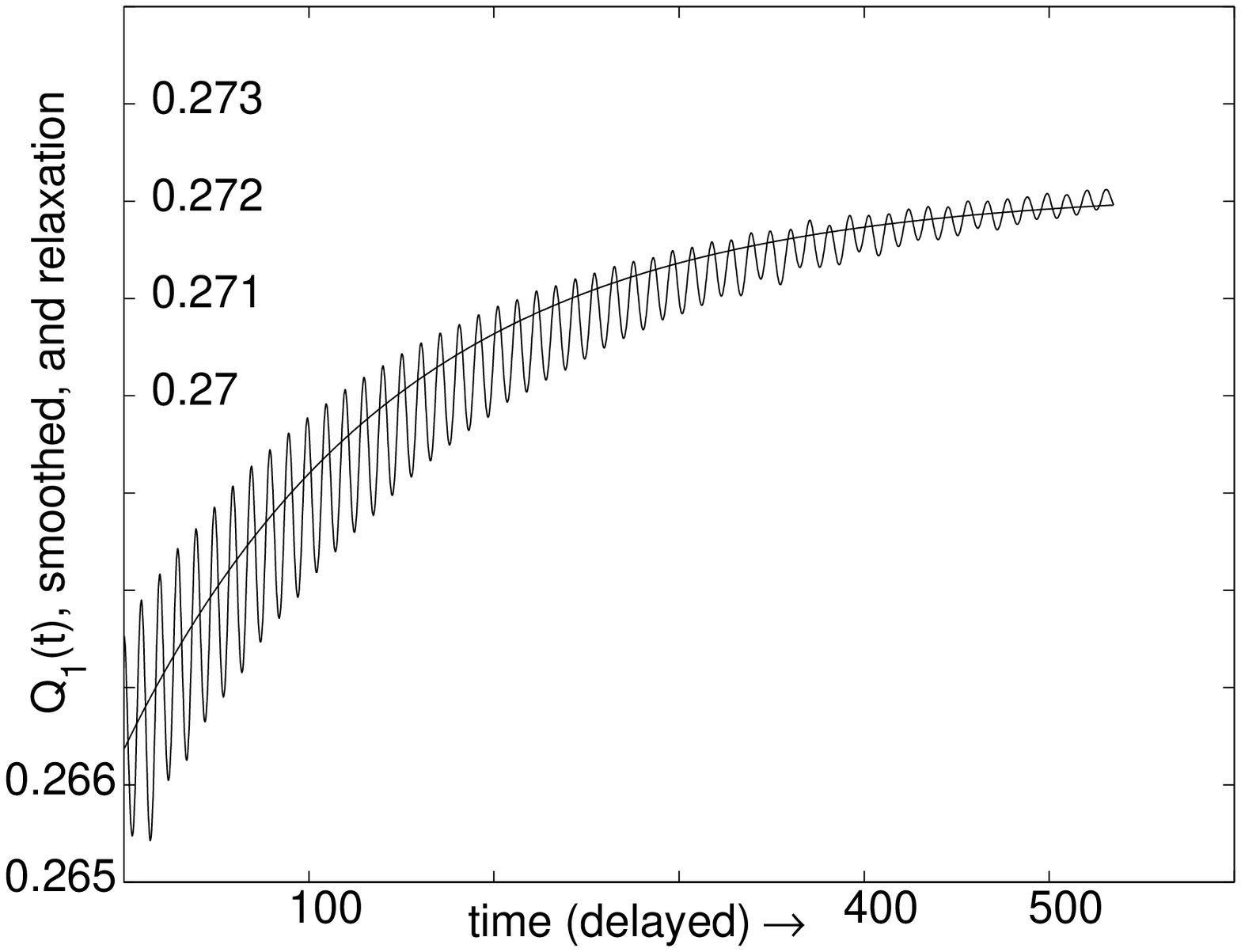,height=\ymeas,width=\xmeas}} \vskip \precap
\caption{The function $Q_1(t)$, partially smoothed, calculated with damping ($\gamma$). Also shown is the curve $1-\exp(-\Gamma t)$ for $\Gamma\approx\gamma\approx0.01$. $Q_1(t)$ is from the confinement run of \figref{histories}.}
\label{expdamp}\end{figure}


\begin{references}
\bibitem{delay} The comparison begins several time steps after the flash, corresponding to the experiment. Picosecond relaxation is not measured and our fits only begin at $\mu$s. In \figref{expdamp} an exact match is not important since our fits \cite{anomalousall} were not sensitive to the exact relaxation function.
\end{references}
\end{document}